\documentclass[preprint,12pt]{elsarticle}

\usepackage{amssymb,amsmath,amsthm,bm}

\biboptions{sort&compress}

\begin{document}

\begin{frontmatter}



\title{New bounds on neutrino electric millicharge
from GEMMA experiment on neutrino magnetic moment}


\author[2]{Victor B. Brudanin}
\ead{brudanin@jinr.ru}
\author[2]{Dmitry V. Medvedev}
\ead{chess1984@mail.ru}
\author[3]{Alexander S. Starostin}
\ead{starostin@itep.ru}
\author[1,2]{Alexander I. Studenikin}
\ead{studenik@srd.sinp.msu.ru}
\address[1]{Department of Theoretical Physics, Faculty of Physics, Moscow State University, Moscow 119991, Russia}
\address[2]{Joint Institute for Nuclear Research, Dubna 141980, Moscow Region, Russia}
\address[3]{Institute for Theoretical and Experimental Physics, National Research Centre "Kurchatovsky Institute", B.Cheremushkinskaya 25, 117218 Moscow, Russia}
\begin{abstract}
Using the new limit on the neutrino anomalous magnetic moment recently obtained by
GEMMA experiment we get an order-of-magnitude estimation for possible new direct
upper bound on the neutrino electric millicharge $\mid q_{\nu} \mid \sim 1.5 \times 10^{-12} e_0$
($e_0$ is the absolute value of the electron charge) by comparing the neutrino magnetic
moment and millicharge contributions to the total cross section at the electron recoil
energy threshold of the experiment. This estimation
is confirmed by the performed analysis of the GEMMA data using established statistical
procedures and a new direct bound on the neutrino millicharge absolute value
$\mid q_{\nu} \mid < 2.7 \times 10^{-12} e_0 \ (90\% CL)$ is derived. This limit is more stringent than the
previous one obtained from the TEXONO reactor experiment data that is included to
the Review of Particle Properties 2012.

\end{abstract}



\end{frontmatter}



\section{Introduction}

The importance of neutrino electromagnetic properties was first mentioned by Wolfgang Pauli just in 1930 when he postulated the existence of this particle and discussed the possibility that the neutrino might have a magnetic moment. Systematic theoretical studies of neutrino electromagnetic properties have started after it was shown that in the extended Standard Model with right-handed neutrinos the magnetic moment of a massive neutrino is, in general, nonvanishing and that its value is determined by the neutrino mass \cite{Fujikawa:1980yx}. In spite of reasonable efforts in studies of neutrino electromagnetic properties, up to now there is no experimental confirmation in favour of nonvanishing neutrino electromagnetic characteristics. The available experimental data in this field do not role out the possibility that neutrinos have ``zero" electromagnetic properties. However, in the course of the recent development of knowledge on neutrino mixing and oscillations, supported by the discovery of flavour conversion of neutrinos from different sources, nontrivial neutrino electromagnetic properties, and nonzero magnetic moment in particular, are straightforward. It is also believed that studies of neutrino electromagnetic properties are important because they provide a kind of bridge (or``open a window") to the new physics beyond the Standard Model. For the recent review on the neutrino electromagnetic properties see \cite{Giunti:2014ixa}.

The neutrino electromagnetic properties are determined by the neutrino electromagnetic vertex
function $\Lambda_{\mu}(q,l)$ that is related to the matrix element of the electromagnetic current between the
neutrino initial $\psi (p)$ and final $\psi (p')$ states. The Lorentz and electromagnetic gauge invariance imply \cite{Kayser:1982br,Nieves:1981zt,
Nowakowski:2004cv,Giunti:2014ixa}
that the electromagnetic vertex function can be written in the form:
\begin{align}
\Lambda_{\mu}(q)
=
f_{Q}(q^{2}) \gamma_{\mu}
-\null & \null
f_{M}(q^{2}) i \sigma_{\mu\nu} q^{\nu} \nonumber
+
f_{E}(q^{2}) \sigma_{\mu\nu} q^{\nu} \gamma_{5}\\
+\null & \null
f_{A}(q^{2}) (q^{2} \gamma_{\mu} - q_{\mu} {q}) \gamma_{5}
,
\label{vert_func}
\end{align}
where where
$f_{Q} $,
$f_{M} $,
$f_{E} $ and
$f_{A} $ are charge, dipole magnetic and electric and anapole neutrino electromagnetic form factors.
Note that the form factors depend only on $q^2$
which is the only independent Lorentz invariant dynamical quantity (the four-vector $q$ is given by $q=p-p'$).

The electric charge is the charge from factor at zero  $q^2$. It is usually believed that
the neutrino has a zero electric charge. This can be attributed to gauge invariance and
anomaly cancelation constraints imposed in the Standard Model. However, if the neutrino
has a mass, the statement that the neutrino electric charge is zero is not so evident as
it meets the eye. In theoretical models with the absence of hypercharge quantization the
electric charge also gets ``dequantized" and, as a result, neutrinos may become electrically
millicharged particles. A detailed discussion of theoretical models predicted the millicharged
neutrinos as well as possible experimental aspects of this problem can be found in many papers
(see, for instance, \cite{Foot:1992ui} and more recent papers \cite{Studenikin:2013my,Studenikin:2012vi, Li:2013pfh,Aisati:2014nda}),
a review on this topic can be found in \cite{Giunti:2014ixa}.

\section{Order-of-magnitude estimation of bound on millicharge}

The strategy of getting constraints on neutrino millicharge from the GEMMA reactor neutrino experiments is as follows \cite{Studenikin:2013my}. Consider a massive neutrino with non-zero electric millicharge $q_{\nu}$ that induces an additional electromagnetic interaction of the neutrino with other particles of the Standard Model.
Such a neutrino behaves as an electrically charged particle
with the direct neutrino-photon interactions, additional to one produced
by possible neutrino non-zero (anomalous) magnetic moment
that is usually attributed to a massive neutrino. If there is no special mechanism of ``screening" of these new electromagnetic interactions then the neutrino will get a normal magnetic moment:
\begin{equation}\label{mu_q}
\mu_{\nu}^{q}=\frac{q_{\nu}}{2m_{\nu}}
\end{equation}
										
Thus, for a millicharged massive neutrino one can expect that the magnetic moment contains two terms,
\begin{equation}\label{1}
\mu_{\nu}^{}=\mu_{\nu}^{q} + \mu_{\nu}^{a}.
\end{equation}
										
Now we consider the direct constraints on the neutrino millicharge obtained using data on the neutrino electromagnetic cross section in the GEMMA experiment. It is important to note that although in the case of a millicharged neutrino two terms, i.e. normal and anomalous magnetic moments, sum up in the total
expression (\ref{1}) for the magnetic moment, however these two contributions should be treated separately when one considers the electromagnetic contribution to the scattering cross section. The point is that the normal magnetic moment contribution is accounted for automatically when one considers the direct neutrino millicharge to the electron charge interaction. The expressions for the neutrino magnetic moment and millicharge cross sections are respectively,
\begin{equation}\label{sigma_mu_1_T}
\left(\frac{d\sigma}{dT}\right)_{\mu^{a}_{\nu}} \approx
\pi\alpha^{2}\frac{1}{m_{e}^{2}T}
\left(\frac{\mu^{a}_{\nu}}{\mu_{B}}\right)^{2},
\end{equation}
and
\begin{equation}\label{sigma_q_e}
\left(\frac{d\sigma}{dT}\right)_{q_{\nu}}\approx 2\pi\alpha
\frac{1}{m_{e}T^2}q_{\nu}^2.
\end{equation}

In case there are no observable deviations from the weak contribution to the neutrino scattering cross section it is possible to get \cite{Studenikin:2013my} the upper bound for the neutrino millicharge demanding that possible effect due
to $q_{\nu}$ does not exceed one due to the neutrino magnetic moment,
\begin{equation}\label{q_limit}
q_{\nu}^{2}<\frac{T}{2m_e}\left(\frac{\mu^{a}_{\nu}}{\mu_{B}}\right)^{2}e_0 ^2.
\end{equation}

Note that the bound derived can be considered as an order-of-magnitude estimation for a possible sensitivity of the experiment to $q_{\nu}$ and a more accurate analysis implies account for experimental data taken over an extended energy range.

\section{New bound on millicharge from GEMMA experiment}

GEMMA (Germanium Experiment for measurement of Magnetic Moment of Antineutrino) investigates the reactor antineutrino-electron scattering at the Kalinin Nuclear Power Plant (Russia). The spectrometer includes a HPGe detector of 1.5 kg installed within NaI active shielding. HPGe + NaI are surrounded with multi-layer passive shielding: electrolytic copper, borated polyethylene and lead. As a result of 4-year measurement the world best upper limit on the neutrino magnetic moments has been obtained \cite{Beda:2012zz}
\begin{equation}\label{mu_bound}
\mu_{\nu}^{a} < 2.9 \times 10^{-11} \mu_{B}.
\end{equation}

Applying this result to (7) and taking into account that the effective threshold T is on the level of 2.8 keV we get \cite{Studenikin:2013my} the upper bound on neutrino millicharge:
\begin{equation}\label{q_2012}
\mid q_{\nu} \mid \sim 1.5 \times 10^{-12} e_0.
\end{equation}
 The obtained constraint should be treated as a rough order-of-magnitude estimation, while the exact values should be evaluated using the corresponding statistical procedures. This is because the limits on the neutrino magnetic moment are derived from the GEMMA experiment data taken over an extended energy range
 from about 2.8 keV to 55 keV rather than at a single electron energy-bin at threshold.

To evaluate the limit on $q_{\nu}$ we use the final spectra from GEMMA. The difference between
$S_{on}$ and $S_{off}$
taking into account $S_{weak}$ normalized by the theoretical electromagnetic spectra $S_{\mu}^{th}$ can
be interpreted as evaluation of $\mu_{\nu}$ and/or $q_{\nu}$ for each energy bin from the region of interest.
The detailed procedure of data processing and obtaining the final result on $\mu_{\nu}^{a}$ is shown in \cite{Beda:2012zz}
and can be illustrated by Fig. 1.
\begin{figure}[h]
\begin{center}
\center{\includegraphics[width=\linewidth]{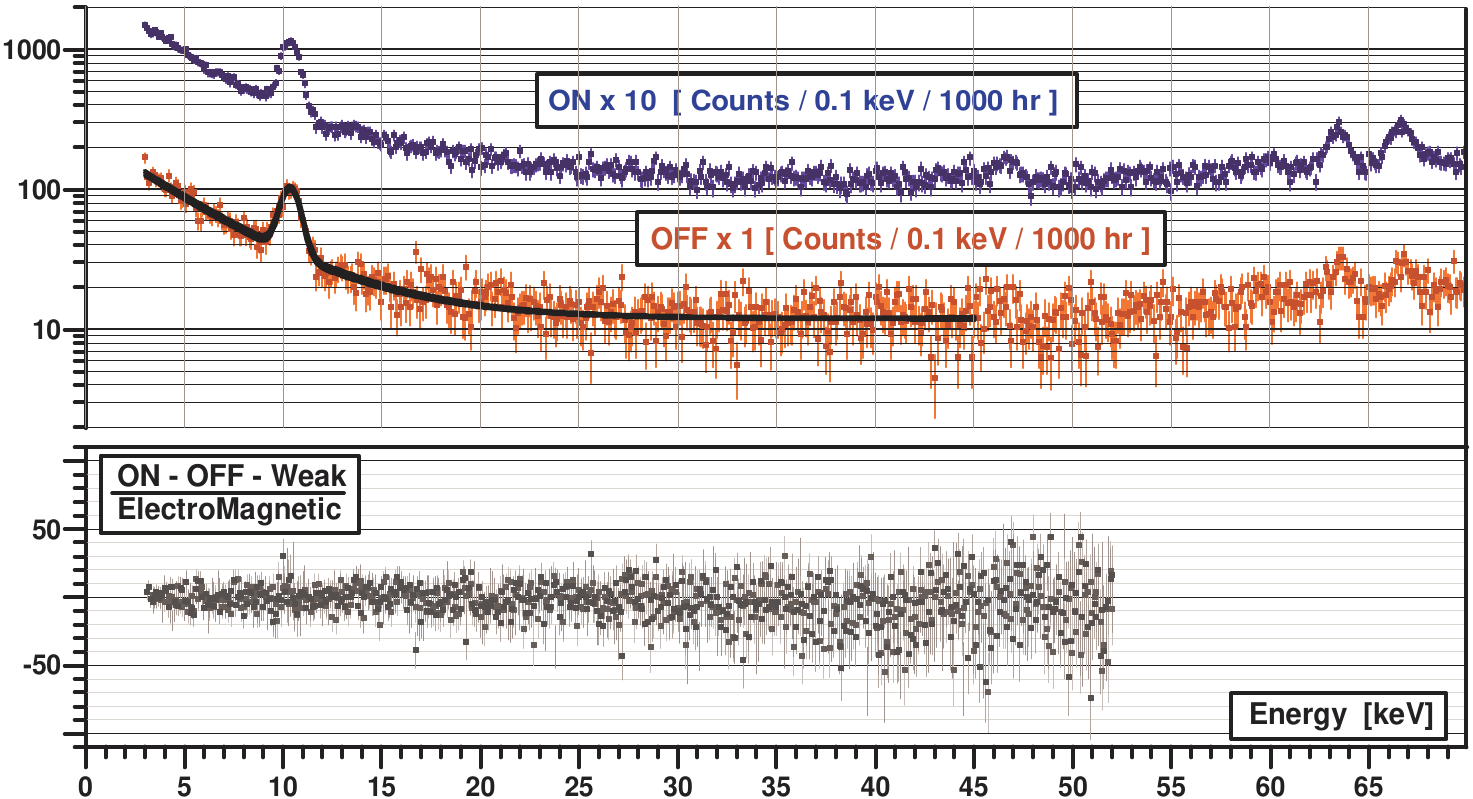}}
\caption{Experimental data of the GEMMA collaboration \cite{Beda:2012zz}.
The difference $S_{on}-S_{off}$ taking into account $S_{weak}$ normalized by the theoretical electromagnetic spectra $S_{\mu}^{th}$.}
\label{figure spectrum}
\end{center}
\end{figure}
The procedure includes the differential method that provides equal pattern for ON and OFF spectra.
The likelihood method is used to obtain the upper limit for an electromagnetic parameter.
Following this procedure we obtain the limit \cite{Studenikin:2013my}
\begin{equation}\label{q_bound_statist_an}
\mid q_{\nu} \mid < 2.7 \times 10^{-12} e_0 \ (90 \% \ C.L.)
\end{equation}
applying the developed procedure to the neutrino millicharge.
										
\section{Future bounds on electromagnetic parameters in GEMMA}

It is interesting to estimate the range of the neutrino millicharge that can be probed in a few years with GEMMA-II experiment that is now in the final stage preparation and is expected to get data in 2015. The experimental setup is being placed under the reactor unit No. 3 where the distance from the centre of the core is 10 m. In this way we double the antineutrino flux up to $5.4\times 10^{13} 1 / cm^2 / s$. Furthermore, being equipped with a special lifting mechanism the spectrometer will be moveable. The mass of the detector is increased by a factor of 4 (two detectors with a mass of 3 kg each). To avoid the``Xe-problems" the internal part of the detector shielding will be gas tight. A special U-type low-background cryostat is used in order to improve the passive shielding and thus reduce the external background in the ROI down to $\sim 0.5 - 1.0 \ (keV\times kg \times day)^{-1}$. A special care is taken to improve antimicrophonic and electric shielding. We also plan to reduce the effective threshold from 2.8 to 1.5 keV.

	As a result GEMMA-II will be sensitive to possible electromagnetic parameters at the level:
\begin{equation}\label{mu_q_bound_II}
\mu_{\nu}^{a} < 1.1 \times 10^{-11} \mu_{B}, \ \ \ \mid q_{\nu} \mid \sim 9.4 \times 10^{-13} e_0.
\end{equation}

For GEMMA-III with new generation detectors ($T_{th} \sim  350 \ eV$) the sensitivity will be
even more improved,

\begin{equation}\label{mu_q_bound_III}
\mu_{\nu}^{a} < 5.8 \times 10^{-12} \mu_{B}, \ \ \ \mid q_{\nu} \mid \sim 5.5 \times 10^{-13} e_0.
\end{equation}	

Note that the obtained estimations for the expected sensitivities of the future GEMMA-II and GEMMA-III to the neutrino millicharge, Eqs. (\ref{mu_q_bound_II}) and (\ref{mu_q_bound_III}), are more conservative than the corresponding sensitivities obtained in \cite {Studenikin:2013my}. This is because the expected sensitivities to $q_{\nu}$ of GEMMA-II and III are derived in \cite {Studenikin:2013my} for a single electron energy-bin at the expected thresholds whereas here above we consider the expected extended energy
ranges for the electron.

\section{Conclusions}

A new upper limit on the neutrino magnetic moment recently obtained by the GEMMA experiment allows us,
by comparing the neutrino magnetic moment and millicharge contributions to the total cross section
at the electron recoil energy threshold of the experiment, to get an order-of-magnitude estimation
for possible new direct upper bound on the neutrino electric millicharge,
	$\mid q_{\nu} \mid \sim 1.5 \times 10^{-12} e_0$. This estimation
is confirmed by performing an analysis of the GEMMA data using established statistical
procedures that yields the new limit at the level of
$\mid q_{\nu} \mid < 2.7 \times 10^{-12} e_0 \ (90 \% \ C.L.)$
The obtained bound (10) on the neutrino millicharge from the recent experimental data of the GEMMA collaboration is more stringent than the reactor neutrino scattering constraint included by the Particle Data Group Collaboration to the Review of Particle Physics \cite{Beringer:1900zz} and that was obtained by \cite{Gninenko:2006fi} from the TEXONO reactor experiment data \cite{Li:2002pn}. Accordingly, we predict that a new bound on the millicharge that can be obtained in future with the new GEMMA experiment data will be a factor of about 10 more stringent than one from the present GEMMA data. Finally, note that upper bounds on the neutrino electric millicharge on the level of $\mid q_{\nu} \mid \sim 10^{-12} e_0$ are also discussed in \cite{Chen:2014dsa}.

\section{Acknowledgments}
One of the authors (A.I.S.) is thankful to Arcadi Santamar\'{\i}a, Salvador Mart\'{\i} and Juan Fuster for the kind invitation to participate in the ICHEP 2014 conference and to all of the organizers for their hospitality in Valencia. This study has been partially supported by the Russian Foundation for Basic Research (grants No. 12-02-00027-a and 14-22-03043-ofi) Russian Scientific Foundation (grant No. 14-12-00920).




\bibliographystyle{model1a-num-names}
\bibliography{1_MedStu_NPB_PS}







\end{document}